\documentstyle[10pt,epsf]{article}
\def\ba{\begin{eqnarray}\samepage}
\def\ea{\end{eqnarray}}
\input amssym.def
\input amssym.tex

\newcommand{\po}{Poincar\'{e} }

\newcommand{\diff}{\partial}

\newcommand{\be}{\begin{equation}}
\newcommand{\ee}{\end{equation}}
\newcommand{\ben}{\begin{eqnarray}\displaystyle}
\newcommand{\een}{\end{eqnarray}}

\def\beq{\begin{equation}}
\def\eeq{\end{equation}}

\begin{document}
\author{Stephen Hewson\footnote{
{\tt Department of Applied Mathematics and Theoretical Physics,
Silver Street,
Cambridge,
CB3 9EW}}\footnote{
 email: {\tt sfh10@damtp.cam.ac.uk}}}
\title{On supergravity in (10,2)}
\maketitle

\begin{abstract}
We consider the problem of creating locally supersymmetric
theories in signature $(10,2)$. The most natural
algebraic starting point is the F-algebra, which is the de Sitter-type
 $(10,2)$
extension of the super-Poincar\'{e} algebra. We derive the corresponding  geometric group
curvatures and evaluate the transformations of the associated gauge fields under the
action of an infinitesimal group element. We then discuss the
formation of locally supersymmetric actions using these quantities. Due to the absence of any
 vielbein terms there is no obvious way to define spacetime as
such. In addition, there is also no way in which we may naturally 
construct an action which is linear in the twelve dimensional curvatures.
We consider the implications of the simplest possible quadratic
theories. We then investigate the relationship between the twelve
dimensional theories and Lorentz signature theories in lower
dimensions. We argue that in this context the process of dimensional
reduction must be replaced by that of group theoretic contraction. Upon
contraction a regular spacetime emerges and we find that
the twelve dimensional curvature constraint reduces to an
Einstein-type equation in which a quadratic non-linearity in the
Ricci scalar is suppressed by a factor of the same magnitude as the
cosmological constant. Finally, we discuss the degrees of freedom of
multi-temporal variables and their relation to ultra-hyperbolic wave
equations. 
	
\end{abstract}
\section {Introduction}
There have been many attempts to construct supersymmetry theories in
twelve dimensions \cite{12}. Although degrees of freedom counting arguments rule
out standard supergravity theories in signatures $(11,1)$ \cite{nahm}, these problems
do not exist in theories with signature $(10,2)$, due to the existence
of Majorana-Weyl spinors \cite{hew1}. How are we to tackle the issue of $(10,2)$
supergravity? Although there have been many efforts in a variety of
directions based upon successful approaches in lower dimensions none
seem to work entirely satisfactorily and often involve a loss of covariance. One way to explain why this
may be the case is that the character of rigid supersymmetry in $(10,2)$
dimensions differs greatly from that in the all important
Lorentzian eleven dimensional scenario. To see why rigid supersymmetry
is so important 
simply note that the basic BPS solutions of the underlying eleven
dimensional superalgebra
correspond exactly to the BPS solutions of the eleven dimensional
supergravity theory \cite{intersect}. Clearly, rigid supersymmetry
theories have a lot to tell us about local supergravities, which is
the viewpoint we adopt in this paper, and since the relevant $(10,2)$ superalgebra
is entirely different in structure  to the eleven dimensional superalgebra 
there is no obvious
reason that a local supersymmetry theory in twelve dimensions should
bear much of a 
resemblance to a traditional supergravity theory. Beginning with the
algebra we present arguments 
as to the restrictions on the possible form of a `supergravity' in twelve
dimensions.

\section{Rigid supersymmetry in eleven and twelve dimensions}
In eleven dimensions rigid supersymmetry is fully described by the
following supersymmetry algebra
\ben
[j_{pq},j_{rs}]&=& 
j_{qs}\eta_{pr} +j_{pr}\eta_{qs}-j_{qr}\eta_{sp}- j_{sp}\eta_{qr} \nonumber \\
\left[j_{pq},p_r\right]&=&p_p\eta_{qr}-p_q\eta_{pr}\nonumber \\
\left[q^{\alpha},j_{pq}\right]&=&\frac{1}{2}{\left(\gamma_{pq}\right)^\alpha}_\beta
q^\beta\nonumber\\
\{q^\alpha,q^\beta\}&=&\left(C\gamma^p\right)^{\alpha\beta}p_p+\frac{1}{2}\left(C\gamma^{pq}\right)^{\alpha\beta}z_{pq}+\frac{1}{5!}(C\gamma^{pqrs\lambda})^{\alpha\beta}z_{pqrs\lambda}\,.
\een
In this algebra the $z$ terms are central and $j$, $p$ and $q$ are the
rotation, momentum and Majorana supersymmetry generators respectively. All
indices are $(10,1)$.  The twelve dimensional
extension of this structure is the F-algebra, which
is a signature $(10,2)$ structure related to the Osp(1,32) group, described in \cite{hew1,vhp,hew2}, in which
$Q^\alpha$ is a 32 component Majorana-Weyl spinor
\ben\label{alegbra}
\left[Q_\alpha,J_{ab}\right]&=&\frac{\Lambda}{2}{(\Gamma_{ab})_\alpha}^\beta
Q_\beta\nonumber\\
\left[Q_\alpha,Z_{abcdef}\right]&=&{\Lambda\over{2}}{(\Gamma_{abcdef})_\alpha}^\beta
Q_\beta\nonumber\\
\left[J_{ab},J_{cd}\right]&=&\Lambda\lbrace
J_{ac}\delta_{bd}+J_{bd}\delta_{ac}-J_{ad}\delta_{bc}-J_{bc}\delta_{ad}\rbrace\nonumber\\
\left[Z_{xbcdef},Z_{ybcdef}\right]&=&\Lambda J_{xy}\nonumber\\
\left[J_{ax},Z_{xbcdef}\right]&=&-\Lambda Z_{abcdef}\nonumber\\
\left[Z_{xyzabc},Z_{xyzdef}\right]&=&\sqrt{\Lambda^2}Z_{abcdef}\\
\left[Z_{xabcde},Z_{xfghij}\right]&=&\Lambda Z_{abcdefghij}=\Lambda
{\epsilon_{abcdefghij}}^{xy}J_{xy}\nonumber\\
\{Q_\alpha,Q_\beta\}&=&\frac{\Delta}{2}\left[\frac{1}{2}(\Gamma^{ab})_{\alpha\beta}J_{ab}+\frac{1}{6!}(\Gamma^{abcdef})_{\alpha\beta}Z_{abcdef}+\frac{1}{10!}(\Gamma^{a_1a_2\dots
a_{10}})_{\alpha\beta}Z_{a_1a_2\dots a_{10}}\right]\nonumber\,,
\een
This algebra is consistent for any choice of the factors $\Lambda$ and
$\Delta$. The general expressions for the $Z_{6}$ commutators are given by
\ben\label{zalgebra}
\left[Z_{a_1\dots a_6},J_{b_1b_2}\right]&=&-6\Lambda\left(\delta_{[a_1|b_1}Z_{b_2|a_2\dots
a_6]}-\delta_{[a_1|b_2}Z_{b_1|a_2\dots
a_6]}\right)\nonumber\\
\left[Z_{a_1\dots a_6},Z_{b_1\dots b_6}\right]&=&-\frac{\sqrt{\Lambda^2}}{120}
\delta^{[a_1a_2a_3}_{[c_1c_2c_3}\delta^{a_4a_5a_6]}_{[b_3b_2b_1}\delta^{c_4c_5c_6]}_{b_4b_5b_6]}{Z^{c_1c_2c_3}}_{c_4c_5c_6}\nonumber\\
&-&36\Lambda\delta^{[a_1}_{[c_1}\delta^{a_2a_3a_4a_5a_6]}_{[b_5b_4b_3b_2b_1}\delta^{c_2]}_{b_6]}{J^{c_1}}_{c_2}\nonumber\\
&-&36\Lambda\delta^{[a_1a_2a_3a_4a_5}_{[c_1c_2c_3c_4c_5}\delta^{a_6]}_{[b_1}\delta^{c_6c_7c_8c_9c_{10}]}_{b_2b_3b_4b_5b_6]}{Z^{c_1c_2c_3c_4c_5}}_{c_6c_7c_8c_9c_{10}}\,, 
\een
with the appropriately chosen pre-factor to obtain the correct
weighting. The expressions involving the ten index $Z$ term are given
by similar formulae, although we treat it as dual to the two index
term $J$ for all purposes except reduction.  
 Note that all the antisymmetrised expressions have
weight one so that, for example, $Z_{[1\dots
6]}=\frac{1}{6!}\left(Z_{123456}-Z_{213456}+\dots\right)$. 
We use the  spinor convention that $\psi_\alpha=\psi^\beta
C_{\beta\alpha}$ and $\psi^\alpha=\psi_\beta C^{\beta\alpha}$. 
The way in which the F-algebra reduces to the eleven dimensional
supersymmetry algebra is via a contraction of the algebra in a
timelike direction, as follows.

\subsection{Contraction of the algebra}
We now perform the operation of contraction on the
theory, in which one of the timelike directions is effectively
decoupled 
 from the system.  
We use the convention the indices $a,b\dots$ run over all $(10,2)$
dimensions, whereas indices $p,q\dots$ take values in  $1\dots 11$. We
define our contraction of the  F-algebra generators as follows
\be
J_{0q}\equiv C_1 p_q\quad J_{pq}\equiv C_2 j_{pq}\quad Z^{0pqrst}\equiv
C_5 z^{pqrst}\quad Z^{pqrstu}\equiv C_6 z^{pqrstu}\,,
\ee
where the $C_i$ are some constants and the lower case letters are to be interpreted as the eleven
dimensional operators. We also write
\be
Q^\alpha\longrightarrow\Theta q^\alpha\,.
\ee 
The contraction process produces a continuum of
consistent deformations, which contain terms qualitatively of the form
\ben\label{deformed}
[z^6,z^6]&\sim&\frac{C_2}{C^2_6}j^2\,,\quad\frac{1}{C_6}z^6\nonumber\\
\left[z^6,z^5\right]&\sim&\frac{1}{C_5}z^6\,,\quad\frac{C_1}{C_5C_6}p^1\nonumber\\
\left[z^5,z^5\right]&\sim&\frac{C_6}{C^2_5}z^6\,,\quad\frac{C_2}{C^2_5}j^2\nonumber\\ 
\left[p^1,z^5\right]&\sim&\frac{C_6}{C_1C_5}z^6\nonumber\\
\left[p^1,z^6\right]&\sim&\frac{C_5}{C_1C_6}z^5\,.
\een
Although any of these contractions are consistent, our aim is  to reproduce the eleven dimensional centrally extended
super-\po algebra, in which case all of the above commutators must
vanish in the infinite radius limit. If we choose to identify the
generators  $j$ with the eleven dimensional Lorentz rotations then the
$[J,J]$ commutator forces us to choose $C_2=\Lambda$. To obtain
additional constraints 
we also look at the anticommutator term
\ben 
\{q,q\}&=&\frac{\Delta}{2\Theta^2}\Big[C_1(\Gamma^t\Gamma^q)p_q+\frac{C_2}{2}(\Gamma^0\Gamma^t\Gamma^{pq})j_{pq}\\
&&\hskip2cm+\frac{C_5}{5!}(\Gamma^t\Gamma^5)z_5+\frac{C_6}{6!}(\Gamma^0\Gamma^t\Gamma^6)z_6
+\frac{C_9}{9!}(\Gamma^t\Gamma^9)z_9+\frac{C_{10}}{10!}(\Gamma^0\Gamma^t\Gamma^{10}z_{10})\Big]\nonumber
\,.
\een
Putting all of this information together we see that in order to
obtain a \po supersymmetric theory we must have the asymptotic relations
\be
C_1+C_{10}\sim\frac{2\Theta^2}{\Delta}\,,\quad C_1\sim\infty\,,\quad C_{10},C_{9},C_6,C_5\sim{\cal{O}}(C_1)\,.
\ee
 With this contraction we obtain an anticommutator of the correct form
\be
\{Q,Q\}=(\gamma^q)p_q+\frac{1}{2}(\gamma^{pq})z_{pq}+\frac{1}{5!}(\gamma^{pqrst})z_{pqrst}\,,
\ee
where, as is usual, we  make use of the algebraic equivalence between
$z^p$ and $z^{11-p}$. Henceforth we shall equate $Z^{10}$ and the dual
of $Z^2$;
$Z^6$ will be self dual. 

\section{Geometry of the supersymmetry theory}
Now that we have evaluated the appropriate rigid supersymmetry in
twelve dimensions let us try to extend our ideas to local
supersymmetry.
Note that the rigid BPS $p$-branes permitted by
the algebra are classified in \cite{hew2}. The basic branes which one
obtains are not $2-$branes and $6-$branes, as one might expect given
2-form and 6-form terms in the algebra and a Lorentzian prejudice, but $2+2$ and $6+2$
dimensional solutions. We also find a natural generalisation of the
$pp$-wave. 
We suppose that local versions of these will provide the fundamental solutions to the local
supersymmetry theory. How are we to find this theory?
Since traditional methods
of supergravity construction fail we look at the geometric
implications of the F-algebra. 
Such geometric ideas have usefully been applied to
supergravities and general relativity in lower dimensions \cite{geometric}.
The first step is to determine the sensible building blocks for
the theory; these are the gauge fields and corresponding curvatures
derived from the algebra. 

\subsection{Curvatures}
Given some coordinate basis on some manifold we define the gauge
covariant derivatives as follows 
\be
\nabla_{\mu}=\diff_\mu -B^{ab}_\mu
J_{ab}-C^{abcdef}_\mu Z_{abcdef}-\psi^\alpha_\mu Q_\alpha\,,
\ee
where 
$\{T_A\}=\{J_{ab},C_{abcdef},Q_\alpha\}$ are the group generators, and
$\{A^A_\mu\}=\{B^{ab}_\mu,C^{abcdef}_\mu,\psi^\alpha_\mu\}$
and the components of the gauge field $A_\mu\equiv A^A_\mu T_A$. The generators $T_A$ of the group will be taken to be in the adjoint representation. The indices $A, B,\dots$ run over the group, whereas the $x^\mu$ are some coordinates on the base manifold ${\cal{M}}$. 
The commutator \footnote{Note that we define the commutator to be
$[X,Y]=XY-YX$, with no factor of one half, whereas antisymmetrised
indices are scaled to have weight one.} of the covariant derivative gives rise to the
curvatures
\ben
[\nabla_\mu,\nabla_\nu]&\equiv&-R^{ab}_{\mu\nu}J_{ab}-R^{(6)}_{\mu\nu}Z_{(6)}-D^\alpha_{\mu\nu}
Q_\alpha\nonumber\\
&=&[\diff_\mu-B^{ab}_\mu J_{ab}-C^{abcdef}_\mu
Z_{abcdef}-\psi^\alpha_\mu Q_\alpha,\diff_\nu-B^{\tilde{a}\tilde{b}}_\nu J_{\tilde{a}\tilde{b}}-C^{\tilde{a}\tilde{b}\tilde{c}\tilde{d}\tilde{e}\tilde{f}}_\nu
Z_{\tilde{a}\tilde{b}\tilde{c}\tilde{d}\tilde{e}\tilde{f}}-\psi^{\tilde{\alpha}}_\nu Q_{\tilde{\alpha}}]\nonumber\\
&=&-2\diff_{[\mu}
B^{ab}_{\nu]}J_{ab}-2\diff_{[\mu}C^{(6)}_{\nu]}Z_{(6)}-2\diff_{[\mu}\psi^\alpha_{\nu]}Q_\alpha\nonumber\\
&+&\Lambda
B^{ab}_\mu B^{cd}_\nu
\left(J_{ac}\delta_{bd}+J_{bd}\delta_{ac}-J_{ad}\delta_{bc}-J_{bc}\delta_{ad}\right)\nonumber\\
&+&2B^{ab}_{[\mu} C^{(6)}_{\nu]}[J_{ab},Z_{(6)}]+2B^{ab}_{[\mu}\psi^{\beta}_{\nu]}[J_{ab},Q_{\beta}]+2C^{(6)}_{[\mu}
B^{ab}_{\nu]}[Z_{(6)},J_{ab}]\nonumber\\
&+&C^{(6)}_\mu
C^{(\bar{6})}_\nu[Z_{(6)},Z_{(\bar{6})}]+2C^{(6)}_{[\mu}\psi^\beta_{\nu]}
[Z_{(6)},Q_\beta]\nonumber\\
&-&\psi^\alpha_\mu\psi^\beta_\nu\{Q_\alpha,Q_\beta\}\,,
\een
as follows
\ben\label{curvatures1}
R^{ab}_{\mu\nu}&=&2\diff_{[\mu}B^{ab}_{\nu]}-4\Lambda B^{c[a}_{[\mu} B^{b]c}_{\nu]}+\Delta\psi^\alpha_\mu\psi^\beta_\nu(\Gamma^{ab})_{\alpha\beta}\nonumber\\
&+&36\Lambda C^{[a|5}_{[\mu}C^{5|b]}_{\nu]}-36\Lambda
C^{Acdefg}_{[\mu}C^{Ahijkl}_{\nu]}\epsilon^{abcdefghijkl}\nonumber\\
D^\alpha_{\mu\nu}&=&2\diff_{[\mu}\psi^\alpha_{\nu]}+{\Lambda}B^{ab}_{[\mu}\psi^\beta_{\nu]}{(\Gamma_{ab})^\alpha}_\beta+{\Lambda}C^{(6)}_{[\mu}\psi^\beta_{\nu]}{(\Gamma_{(6)})^\alpha}_\beta\nonumber\\
R^{abcdef}_{\mu\nu}&=&2\diff_{[\mu}C^{abcdef}_{\nu]}+\frac{\sqrt{\delta^2}}{120}C^{ABC[abc}_{[\mu}C^{def]ABC}_{\nu]}\nonumber\\
&-&24\Lambda
B^{aA}_{[\mu}C^{Abcdef}_{\nu]}+\frac{\Delta}{6!}\psi^\alpha_\mu\psi^\beta_{\nu}(\Gamma^{abcdef})_{\alpha\beta}\,.
\een
Finally, we note that the Jacobi identity for the curvatures arising
from
\be
[\nabla_{(\mu},[\nabla_{\nu},\nabla_{\rho)}]]=0\,,
\ee
where round brackets indicate that we sum over cyclic permutations,
lead to the constraints
\be
\diff_{(\mu} R^A_{\rho\sigma)}-h^B_{(\mu}R^C_{\rho\sigma)}{f_{AB}}^A=0\,.
\ee
\subsection{Transformation of the curvatures and gauge fields}

Under the action of an infinitesimal group element
$S=\exp(\epsilon\cdot T), \epsilon=(\eta,\omega,\Omega)$, the gauge field
transforms as
\be
A'_\mu=SA_\mu S^{-1} + (\diff_\mu S)S^{-1}\,,
\ee
whereas the curvatures transform homogeneously as
\be
R'_{\mu\nu}=SR'_\mu S^{-1}\,.
\ee
Explicitly, to first order in the parameter $\epsilon$, we find that 
\ben
A'_\mu&=&(\diff_\mu\eta)\cdot Q+(\diff_\mu\omega)\cdot J +
(\diff_\mu\Omega)\cdot Z\nonumber\\
&+&\hskip-0.3cm (1+\eta\cdot Q+ \omega\cdot J +\Omega \cdot Z)\Big(B^{ab}_\mu
J_{ab}+\psi^\alpha_\mu Q_\alpha+C^{(6)}_\mu Z_{(6)}\Big)(1-\eta\cdot
Q- \omega\cdot J -\Omega \cdot Z)\,.
\een
giving us
\ben
\delta A_\mu&=&(\diff_{\mu}\eta)\cdot Q+(\diff_\mu\omega)\cdot J +
(\diff_\mu\Omega)\cdot Z\nonumber\\
&\ &\hskip1cm +\left[\eta\cdot Q+\omega\cdot J+\Omega\cdot Z,B^{ab}_\mu
J_{ab}+\psi^\alpha_\mu Q_\alpha+C^{(6)}_\mu Z_{(6)}\right]\nonumber\\
&=&(\diff_{mu}\eta)\cdot Q+(\diff_\mu\omega)\cdot J +
(\diff_\mu\Omega)\cdot
Z-\eta^\alpha\psi^\beta_\mu\{Q_\alpha,Q_\beta\}\nonumber\\
&\ &+\omega^{ab}B^{cd}_{\mu}[J_{ab},J_{cd}]+\Omega^{(6)}C^{(\bar{6})}[Z_{(6)},Z_{(\bar{6})}]+\eta^\alpha B^{ab}_\mu[Q_\alpha,J_{ab}]+\eta^\alpha
C^{(6)}_{\mu}[Q_\alpha,Z_{(6)}]\nonumber\\
&\ &\hskip1cm +\omega^{ab}\psi^\alpha_\mu[J_{ab},Q_\alpha]+\omega^{ab}C^{(6)}_\mu[J_{ab},Z_{(6)}]+
\Omega^{(6)}B^{ab}_\mu[Z_{(6)},J_{ab}]+\Omega^{(6)}\psi^\alpha_\mu[Z_{(6)},Q_\alpha]\nonumber\\
&\equiv& \delta B^{ab}_\mu J_{ab}+\delta\psi^{\alpha}_\mu Q_\alpha
+\delta C^{(6)}_\mu Z_{(6)}\,.
\een
Expanding the anticommutators gives us the resulting changes in the
gauge potentials under a gauge transformation
\ben
\delta\psi^\alpha_\mu&=&(\diff_\mu\eta)^\alpha+\frac{\Lambda}{2}\eta^\beta
B^{ab}_\mu(\Gamma_{ab})^{\alpha}_\beta+\frac{\Lambda}{2}\eta^\beta
C^{(6)}_\mu(\Gamma_{(6)})^\alpha_\beta-\frac{\Lambda}{2}\omega^{ab}\psi^\beta_\mu(\Gamma_{ab})^\alpha_\beta-\frac{\Lambda}{2}\Omega^{(6)}\psi^\beta_\mu(\Gamma_{(6)})^\alpha_\beta\nonumber\\
\delta
B^{ab}_\mu&=&(\diff_\mu\omega^{ab})-\Delta\eta^\alpha\psi^\beta_\mu(\Gamma^{ab})_{\alpha\beta}-4\Lambda\omega^{ac}B^{cb}_\mu-36\Lambda\Omega^{a5}C^{b5}_\mu+36\Lambda\Omega^{A5}C^{A\bar{5}}_\mu\epsilon_{ab5\bar{5}}\\
\delta C^{abcdef}_\mu&=&(\diff_\mu\Omega^{(6)})+\frac{\sqrt{\Lambda^2}}{120}
\Omega^{ABC[abc}C^{def]ABC}_\mu
-\frac{\Delta}{6!}\eta^\alpha\psi^\beta_\mu(\Gamma^{(6)})_{\alpha\beta}\nonumber\\
&&\hskip2cm+12\Lambda\omega^{aA}C^{Abcdef}_\mu-12\Lambda
B^{aA}_\mu\Omega^{Abcdef}\nonumber\,.
\een
There are thus three different types of gauge transformation
associated with the F-algebra. We now detail the effects of these on
the curvatures in turn
\begin{enumerate}
\item
{\bf ${\epsilon=(0,\omega,0)}$}

If $\eta=0$ and $\Omega=0$ then the gauge potentials transform as
\ben
\delta\psi^\alpha_\mu&=&-\frac{\Lambda}{2}\omega^{ab}\psi^\beta_\mu{(\gamma_{ab})^\alpha}_\beta\nonumber\\
\delta B^{ab}_\mu&=&\diff_\mu\omega^{ab}+4\Lambda\omega^{ac}B^{cb}_\mu\nonumber\\
\delta C^{abcdef}_\mu&=&12\Lambda\omega^{aA}C^{Abcdef}_\mu\,,
\een
and the curvature terms transform as under a Lorentz transformation
\ben
\delta R^{ab}_{\mu\nu}&=&-4\Lambda\omega^{ac}R^{cb}_{\mu\nu}\nonumber\\
\delta R^{abcdef}_{\mu\nu}&=&-12\Lambda\omega^{aA}R^{Abcdef}_{\mu\nu}\nonumber\\
\delta D^\alpha_{\mu\nu}&=&-\frac{\Lambda}{2}\omega^{ab}{(\Gamma_{ab})^\alpha}_\beta D^\beta_{\mu\nu}\,.
\een

\item
{\bf ${\epsilon=(0,0,\Omega)}$}

In this situation, the gauge potentials transform as 
\ben
\delta\psi^\alpha_\mu&=&-\frac{\Lambda}{2}\Omega^{(6)}\psi^\beta_\mu(\Gamma_{(6)})^\alpha_\beta\nonumber\\
\delta B^{ab}_\mu&=&-36\Lambda\Omega^{a5}C^{b5}_\mu+36\Lambda\Omega^{A5}C^{A\bar{5}}_\mu\epsilon_{ab5\bar{5}}\nonumber\\
\delta C^{abcdef}_\mu&=&
\diff_\mu\Omega^{(6)}+\frac{\sqrt{\Lambda^2}}{120}
\Omega^{ABC[abc}C^{def]ABC}_\mu -12\Lambda\Omega^{bcedfA}B^{Aa}_\mu\,,
\een
giving rise to a 6-form version of a Lorentz transformation as follows
\ben
\delta\psi^\alpha_{\mu\nu}&=&-\frac{\Lambda}{2}\Omega^{(6)}{(\Gamma_{(6)})^\alpha}_\beta
D^\beta_{\mu\nu}\nonumber\\
\delta R^{ab}_{\mu\nu}&=&36\Lambda\Omega^{a5}R^{b5}_\mu+36\Lambda\Omega^{A5}R^{A\bar{5}}_\mu\epsilon_{ab5\bar{5}}\nonumber\\
\delta R^{abcdef}_{\mu\nu}&=&\frac{\sqrt{\Lambda^2}}{120}
\Omega^{ABC[abc}R^{def]ABC}_\mu-12\Lambda\Omega^{bcedfA}R^{Aa}_\mu\,,
\een

\item
{\bf ${\epsilon=(\eta,0,0)}$}

A fermionic gauge transformation leads us to 
\ben\label{susytrans}
\delta\psi^\alpha_\mu&=&\diff_\mu\eta^\alpha+\frac{\Lambda}{2}\eta^\beta
B^{ab}_\mu(\Gamma_{ab})^{\alpha}_\beta+\frac{\Lambda}{2}\eta^\beta
C^{(6)}_\mu(\Gamma_{(6)})^\alpha_\beta\nonumber\\
\delta
B^{ab}_\mu&=&-\Delta\eta^\alpha\psi^\beta_\mu(\Gamma^{ab})_{\alpha\beta}\nonumber\\
\delta C^{abcdef}_\mu&=&
-\frac{\Delta}{6!}\eta^\alpha\psi^\beta_\mu(\Gamma^{abcdef})_{\alpha\beta}\,,
\een
We shall call such a variation a {\it supersymmetry transformation}. These
act on the curvatures to give
\ben
\delta
D_{\mu\nu\alpha}&=&\frac{\Lambda}{2}\big(R^{ab}_{\mu\nu}(\Gamma_{ab})_{\alpha\gamma}+R^{6}_{\mu\nu}(\Gamma_{6})_{\alpha\gamma}\big)\eta^\alpha\nonumber\\
\delta R^{ab}_{\mu\nu}&=&-\Delta\eta^\alpha
D^\beta_{\mu\nu}(\Gamma^{ab})_{\alpha\beta}\nonumber\\
\delta R^6_{\mu\nu}&=&-\frac{\Delta}{6!}\eta^\alpha
D^\beta_{\mu\nu}(\Gamma^6)_{\alpha\beta}\,.
\een

\end{enumerate}
We now discuss the construction of Lagrangians from the curvatures and
potentials which give rise to invariant actions.

\section{Supergravities, supersymmetry and Lagrangians}
The goal is to construct some locally symmetric supergravity-type
theory in signature $(10,2)$. As a basic starting point we suppose that
we must use the potentials and curvatures to create a scalar
Lagrangian which gives rise to an invariant action 
\be
S=\int {\cal{L}}\,.
\ee
Clearly the action is to  be constructed
from elements with indices of type $\mu$ and $A$. The fact that we
require a {\it scalar} Lagrangian presents us with an interesting
problem: how can we create a scalar using our basic fields?  The
obvious way to contract the group indices is via the {\it Killing
form} $B$ of our Lie superalgebra, defined  as follows
\be\label{killingform}
B(a,b)=\mbox{str}\big(ad({\bf a})ad({\bf b})\big)\,,
\ee
for elements ${\bf a, b}$ of the Lie algebra, where `str' denotes the operation of supertrace.  The Killing form
essentially provides us with a metric on the superalgebra, the
components of which are given by
\be\label{killingmetric}
{\cal{G}}_{AB}=B(T_A,T_B)={f^{D}_{AC}}{f^{C}_{BD}}\,,
\ee
where $T_A$ are the generators of the algebra. For the F-algebra, a
calculation shows that the only non-zero metric elements are given by
\ben
{\cal{G}}_{\alpha\beta}&=&\frac{2\Delta}{\Lambda}C_{\alpha\beta}\nonumber\\
{\cal{G}}_{[a][b]}&=&
\delta_{a_1b_1}\delta_{a_2b_2}-\delta_{a_1b_2}\delta_{a_2b_1}\nonumber\\
{\cal{G}}_{[A][B]}&=& 6!\left(\delta^{A_1\dots A_6}_{B_1\dots
B_6}+{\epsilon_{A_1\dots A_6B_1\dots B_6}}\right)\,,
\een
in an obvious notation. 
Whereas we may naturally
contract the group indices $A$ with the assistance of our new metric
${\cal{G}}_{AB}$ and its inverse, in order to contract on the indices
$\mu, \nu\dots$ we must introduce an additional `metric' term into our
theory. In
ordinary gauge theories of supergravity, which are based on \po groups,
there is a natural choice for the metric, since the momentum
generators $p_\mu$ give rise to a potential term $e^A_\mu$, which can
be treated as the vielbein. This enables us to relate the metric
$g_{\mu\nu}$ on the base to the group space metric as follows
\be\label{vielbein}
e^A_\mu e^B_\nu{\cal{G}}_{AB}=g_{\mu\nu}\,.
\ee
Unfortunately, for the F-algebra there are no momentum-type
generators  and we cannot, therefore,  define a spacetime metric in
this way whilst maintaining covariance.  However, after contracting
the F-algebra, we saw that we can recover the eleven dimensional 
superalgebra. This, of course, contains momentum generators which may
be used to generate a proper metric term.  We therefore anticipate that
there does not exist a covariant $(10,2)$ supergravity in the usual
sense of the word, but some other geometric theory, upon {\it
contraction} of which a
supergravity emerges.

\subsection{Actions}

We should now consider the possible forms of the action available to
us. Although Einstein gravity suggests that we write down a form of
the action which is linear in the curvatures, without a vielbein term
is is not possible to write down such an action in a natural fashion
because there is no way to relate the group indices $A$ to the base
indices $\mu$. We therefore look for actions which are quadratic in
the group curvatures, the simplest choice being of Yang--Mills type
\be\label{lagrangian} 
{\cal{L}}=\sqrt{g}R^A_{\mu\nu}R^{B}_{\rho\sigma}g^{\mu\rho}g^{\nu\sigma}{\cal{G}}_{AB}\,,
\ee
where $g$ is some  metric on the base $M$, $g$ is
its determinant and ${\cal{G}}_{AB}$ is the metric derived from the Killing form on the
group. Henceforth we shall raise and lower all Greek indices with the
supposed 
metric $g_{\mu\nu}$, the inverse of which is to be  defined through the usual
relationship
\be
g^{\mu\sigma}g_{\sigma\nu}={1^\mu}_{\nu}\,.
\ee
Latin indices are raised and lowered with the use of the group space
metric ${\cal{G}}_{AB}$. 
Since the Killing metric is constructed as a supertrace of the group
generators, and because the curvatures transform homogeneously under a
gauge transformation, both of these actions are invariant under gauge
transformations. 
The equations of motion for each of the gauge fields are as follows\footnote{We assume that $g$ is not a function of derivative terms.} 
\ben
\frac{1}{\sqrt{g}}\diff_\nu(\sqrt{g}D_{\alpha\mu\nu})&=&\Delta D^\gamma_{\mu\nu}\left(B^{ab}_\nu(\Gamma_{ab})_{\gamma\alpha}+C^6_\nu{(\Gamma_6)}_{\gamma\alpha}\right)+\Delta\psi^\beta_\mu\left(R^{ab}_{\mu\nu}(\Gamma_{ab})_{\alpha\beta}+R^{(6)}_{\mu\nu}(\Gamma_{(6)})_{\alpha\beta}\right)
+E(D)_{\alpha\mu}\nonumber\\
\frac{1}{\sqrt{g}}\diff_\nu(\sqrt{g}R^{ab}_{\mu\nu})&=&-\Delta
D^\alpha_{\mu\nu}\psi^\beta_\nu{(\Gamma^{ab})}_{\alpha\beta}
-24\Lambda R^{a5}_{\mu\nu} C^{b5}_\nu+8\Lambda
B^{bc}_{\nu}R^{ca}_{\mu\nu}+E(R)^{ab}_\mu\nonumber\\
\frac{1}{\sqrt{g}}\diff_\nu(\sqrt{g}R^6_{\mu\nu})&=&-\frac{\Delta}{6!}
D^\alpha_{\mu\nu}\psi^\beta_\nu (\Gamma_6)_{\alpha\beta}- 24\Lambda
R^{a5}_{\mu\nu}
B^{1a}_\nu+\frac{2\sqrt{\delta^2}}{120}R^{3abc}_{\mu\nu}C^{\bar{3}abc}_\nu+E(R)^{6}_\mu\,,
\een
where 
\ben
E(D)_{\alpha\mu}&=&\frac{\diff(\sqrt{g}g^{\mu\rho}g^{\nu\sigma})}{\diff
\psi^\alpha_\nu}R^A_{\mu\nu}R^{B}_{\rho\sigma}{\cal{G}}_{AB}\nonumber\\
E(R)^{ab}_{\mu}&=&\frac{\diff(\sqrt{g}g^{\mu\rho}g^{\nu\sigma})}{\diff
B^{ab}_\nu}R^A_{\mu\nu}R^{B}_{\rho\sigma}{\cal{G}}_{AB}\nonumber\\
E(R)^{6}_{\mu}&=&\frac{\diff(\sqrt{g}g^{\mu\rho}g^{\nu\sigma})}{\diff
C^{6}_\nu}R^A_{\mu\nu}R^{B}_{\rho\sigma}{\cal{G}}_{AB}\,.
\een

\subsection{Supersymmetry transformations}
We now vary the Lagrangian with the {\it supersymmetry
transformations} (\ref{susytrans}). Using the fact that the
combination $\sqrt{g}
g^{\mu\rho}g^{\nu\sigma}$ transforms as 
\be
\delta\left(\sqrt{g}
g^{\mu\rho}g^{\nu\sigma}\right)=\sqrt{g}\left(\frac{1}{2}\delta
g_{\lambda\delta} g^{\lambda\delta}g^{\mu\rho}g^{\nu\sigma}
-\delta
g_{\tilde{\rho}\tilde{\sigma}}g^{\mu\tilde{\rho}}g^{\rho\tilde{\sigma}}g^{\nu\sigma}
-\delta
g_{\tilde{\rho}\tilde{\sigma}}g^{\nu\tilde{\rho}}g^{\sigma\tilde{\sigma}}g^{\mu\rho}
\right)
\ee
we find that the overall variation of the action, after much
simplification, is just proportional to the variation of the metric
under supersymmetry, for any choice of $g_{\mu\nu}$
\be
\delta{\cal{L}}=\delta
g_{\mu\nu}\left(-2{{R^\mu}_\sigma}^A{R^{\nu\sigma}_A}+\frac{1}{2}g^{\mu\nu}R^{A\rho\sigma}R_{A\rho\sigma}\right)
\ee

\subsection{The metric terms}
Let us now consider various possibilities for the metric field, and
the implications for the supersymmetry of the theory. There are
essentially two  basic choices  for the form of $g_{\mu\nu}$:
Either it is a function of the gauge fields, in which case there is
some arbitrariness concerning the exact choice of the function,  or it is some additional
independent field. We discuss these scenarios in turn

\begin{enumerate}

\item

Firstly  we consider the case in which the metric is an independent
field, in which case  $E(D)=E(R)=0$. For this choice of $g_{\mu\nu}$ the local
supersymmetry variation of the entire action vanishes. Furthermore, we
obtain an 
additional equation of  motion from the variation of the action with
respect to the field $g$, similar to a `quadratic Einstein equation'
\be
\frac{1}{4}g_{\mu\nu}R^A_{\rho\sigma}R^{\rho\sigma}_A=R^A_{\mu\rho}{R_{A\nu}}^\rho\,.
\ee
Thus the Lagrangian is on-shell supersymmetric. Let us look at the
equation of motion implied by the metric term: taking the trace implies that the squared curvature
vanishes
\be
R^2\equiv R^{A\mu\nu}R_{A\mu\nu}=0\,,
\ee
which implies that the $g$ equation of motion becomes
\be
R^A_{\mu\rho}{R_{A\nu}}^\rho=0\,.
\ee
\item
If we
are to construct the metric from fields in the problem then there is one
natural possibility: use the metric obtained with the use of the Killing
form
\be     
\hskip1cm g_{\mu\nu}=\frac{2\Delta}{\Lambda}\psi^\alpha_\mu\psi^\beta_\nu C_{\alpha\beta}+B^{ab}_\mu B_{\nu ab}+6!C^{(6)}_\mu
C_{\nu(6)}\,,
\ee
which gives us a Born-Infeld style theory. 
In this case we find that
\ben\label{firstE}
E(D)_{\mu\alpha}&=&\frac{2\Delta}{\Lambda}\psi_{\mu\alpha}\left(-2{{R^\mu}_\sigma}^A{R^{\nu\sigma}}_A+\frac{1}{2}g^{\mu\nu}{R^{\mu\nu}_AR^A_{\mu\nu}}\right)\nonumber\\
E(R)^{ab}_\mu&=&B^{ab}_\mu
\left(-2{{R^\mu}_\sigma}^A{R^{\nu\sigma}_A}+\frac{1}{2}g^{\mu\nu}{R^{\mu\nu}_AR^A_{\mu\nu}}\right)\nonumber\\
E(R)^{6}_\mu&=&C^{(6)}_\mu
\left(-2{{R^\mu}_\sigma}^A{R^{\nu\sigma}_A}+\frac{1}{2}g^{\mu\nu}{R^{\mu\nu}_AR^A_{\mu\nu}}\right)\,,
\een
whereas the metric $g_{\mu\nu}$ has the very simple variation  
\be
\delta
g_{\mu\nu}=\frac{4\Delta}{\Lambda}(\diff_\mu\eta^\alpha)C_{\alpha\beta}\psi^\beta_\nu\,.
\ee
In this scenario, the overall  variation of the Lagrangian under
supersymmetry is given by
\be\label{variation1}
\delta{\cal{L}}=-2\frac{\Delta}{\Lambda}\diff\eta^\alpha_\mu\psi_{\nu\alpha}\left(4{{R^\mu}_\sigma}^A{R^{\nu\sigma}_A}-g^{\mu\nu}R^{A\rho\sigma}R_{A\rho\sigma}\right){\cal{G}}_{AB}\,,
\ee
This is an interesting situation: we see that if the SUSY parameter
$\eta$ is constant then the variation is identically zero. Another
point is that the action is identically off-shell supersymmetric if we
work with a special class of spinors which lie on a `quadratic cone' such that 
\be
C_{\alpha\beta}\psi^\alpha\phi^\beta=0\,,
\ee
for any pair of spinors $\phi$ and $\psi$. This expression is
equivalent to the statement that the spinors would be Majorana-Weyl in
signature $(9,1)$. In the $(10,2)$ signature it merely defines a new class
of spinors\footnote{A discussion of the relationship between these spinors and
Dirac spinors is given in \cite{hew1}}. If, however,  we do not wish to impose restrictions on the type of
spinors in the problem then the variation of the action becomes
 zero if we require that  the curvatures obey the equation 
$
\frac{1}{4}g_{\mu\nu}R^A_{\rho\sigma}R^{\rho\sigma}_A=R^A_{\mu\rho}{R_{A\nu}}^\rho\,,
$ which is something akin to a quadratic form of the Einstein
equations. Taking this trace of the equation again implies a
Ricci-flat type condition
\be
R^A_{\mu\rho}{R_{A\nu}}^\rho=0\,.
\ee
\end{enumerate}

\section{Reduction of multi-temporal supergravity}
The main idea of this paper is that the locally supersymmetric theory
in signature $(10,2)$ should reduce down to a known supergravity by the
process of contraction in one of the timelike group directions, as opposed to dimensional
reduction of an underlying spacetime structure. This idea is supported
by the fact that ordinary compactification does not work in quite them
same way in a multi-temporal space. The reason for this is that one
must compactify on a Lorentzian internal space. From a very general
algebraic viewpoint a compactification which preserves a supersymmetry
requires the internal space to be a spin manifold with special
holonomy. In $M$-theory, which is eleven dimensional, these holonomy groups are $SU(2), SU(3), G_2$
and $Spin(7)$, corresponding to manifolds of dimension $4,6,7,8$
respectively \cite{holonomy}.
To compactify a $(10,2)$ theory to a Lorentzian theory in eleven
dimensions or less would require us to reduce on a compact spin manifold of
dimension $(n,1)$ where $n\leq 9$. There are no such irreducible
manifolds \cite{nolorentz}. Thus,
reduction will simply not work in the same way as is usual, which  gives us the confidence that one must abandon the idea of
dimensional reduction and instead resort to algebraic contraction. 
In this context is is
natural to provide a new interpretation to  quantities which involve a
{\it group} index corresponding to the contraction direction, as follows
\be
B^{0p}=e^p\quad\quad C^{0pqrst}=e^{pqrst}
\ee
In particular we have the natural emergence of a vielbein term.  Furthermore,
the form of the curvatures after reduction certainly gives us the
possibility of 
constructing an ordinary supergravity in lower dimensions. In the
scaled limit in which many of the terms in the algebra become central we
find the following effectively eleven dimensional quantities
\ben
R^{pq}_{\mu\nu}&=&\frac{2}{\Lambda}\left(\diff_{[\mu}B^{pq}_{\nu]}-2
B^{r[p}_{[\mu}
B^{q]r}_{\nu]}\right)+\Lambda e^{[p}_{[\mu} e^{q]}_{\nu]}\nonumber\\
R^{0p}_{\mu\nu}&=&4\diff_{[\mu}e^p_{\nu]}-4e^q_{[\mu}B^{pq}_{\nu]}+\psi^\alpha_\mu\psi^\beta_\nu(\Gamma^{0p})_{\alpha\beta}\nonumber\\
D^\alpha_{\mu\nu}&=&2\diff_{[\mu}\psi^\alpha_{\nu]}+B^{ab}_{[\mu}\psi^\beta_{\nu]}{(\Gamma_{ab})^\alpha}_\beta\nonumber\\
R^{pqrstu}_{\mu\nu}&=&2\diff_{[\mu}C^{pqrstu}_{\nu]}+24
B^{pv}_{[\mu}C^{vqrstu}_{\nu]}+\psi^\alpha_{[\mu}\psi^\beta_{\nu]}\frac{1}{6!}(\Gamma^{pqrstu})_{\alpha\beta}\nonumber\\
R^{0pqrst}_{\mu\nu}&=&2\diff_{[\mu}e^{pqrst}_{\nu]}+24
B^{up}_{[\mu}e^{0uqrst}_{\nu]}+\psi^\alpha_{[\mu}\psi^\beta_{\nu]}\frac{1}{6!}(\Gamma^{0pqrst})_{\alpha\beta}
\,.
\een 
Let us investigate the interpretation of these equations. 
If we treat the $B$ as a spin connection then we se familiar objects emerge
\ben
\diff_{[\mu}B^{pq}_{\nu]}-2
B^{r[p}_{[\mu}
B^{q]r}_{\nu]}&\longrightarrow& \mbox{Riemann curvature}\nonumber\\
\diff_{[\mu}e^p_{\nu]}-e^q_{[\mu}B^{pq}_{\nu]}&\longrightarrow&\mbox{Torsion}\nonumber\\
D^\alpha_{\mu\nu}&\longrightarrow&\nabla_{{[}\mu}\psi^\alpha_{\nu]}
\een
What of the $Z$ terms? 
Although the two index and six index objects are originally treated
equally by the group theory, after the contraction the two index term
becomes a Lorentz rotation $j$, whereas all other terms become central and
commute with everything except $j$. Let us look at the way in which
the curvature associated with the Lorentz rotations behaves after
contraction. Setting the  fermion and six-form terms to zero we see
that our twelve
dimensional equation of motion $R^A_{\mu\rho}{R_{A\nu}}^\rho=0$
reduces to the following
\be
R+\Lambda=-\frac{R^2}{\Lambda}\,,
\ee
where $R$ is an effective eleven dimensional Ricci scalar. Although 
this equation is non-linear in $R$, we see that the non-linearity is
of a size inversely proportional to the magnitude of a cosmological
constant. Thus a very large cosmological constant implies that the
theory reduces to an only  very
slightly perturbed linear equation. This is very important: if the
non-linearities in the Riemann tensor were not suppressed that there
would be no way in which to relate the theory to traditional gravity
theories in lower dimensions. Inclusion of the other terms in
the very large $\Lambda$ limit provides us with an equation
qualitatively of the form
\be
R+|\nabla F_7|^2+|\nabla\phi|^2+\mbox{fermi}+\Lambda=0
\ee
where $F_7$ is a seven-form term derived from $C^{6}_\mu$.

\subsection{Degrees of freedom}
Of course, in any study of supergravity we are crucially interested in
the degrees of freedom of the fields in question. 
What does this mean in the context of a theory with two timelike
directions? Essentially, the number of degrees of freedom of a given
variable is determined by the amount of data needed to specify uniquely
the solution to the underlying equation of motion. Physical one-time
quantities propagate via wave equations, which are well understood: In order to specify the
solution to an $n$-dimensional wave equation uniquely we must give
data on some $(n-1)$-dimensional Cauchy surface, which evolve along
light cones. In order to specify the degrees of freedom of an $(n+2,2)$
dimensional variable we would need to understand the way in which we
can solve {\it ultra-hyperbolic wave equations} 
\be
\nabla^2_n\phi=\frac{\partial^2\phi}{\partial
t^2_1}+\frac{\partial^2\phi}{\partial t^2_2}\,,
\ee
where $t_1$ and $t_2$ are the two timelike directions. 
Mathematicians are only now beginning to study in depth the properties
of multi-temporal wave equations \cite{brenner}. 
Unfortunately,
it is not known how to fully generalise the Cauchy problem to
multi-temporal scenarios, which is a significant stumbling
block to the understanding of the role of many times in
M-theory. Although one may show that Cauchy problems or
boundary value problems are  ill
posed in the ultra-hyperbolic scenario, this does not mean to say that
there is not some $X$-problem for which the solution to an
ultra-hyperbolic partial differential equation is always uniquely
determined and bounded. An understanding of this problem would shed enormous
insight into the possible role of many times in physics. 
However, one resolution may be to prescribe data on a null cone,
which then itself evolves along a null cone. 
Naive counting along these lines would suggest that a vector index would be
described by the `double light cone'. Thus a vector index would have 
$D-4$ degrees of freedom, corresponding to the number of directions
orthogonal to the `double lightcone'

\section{Conclusion}
We have discussed some of the problems associated with the formulation of a
supergravity theory in twelve dimensions. Although one may always
speculate on the form of such a theory, a safe base from which
to begin the exploration is the $(10,2)$ signature F-algebra, because this
contracts to produce the eleven dimensional supersymmetry algebra.
 Since this twelve dimensional algebra does not contain any momentum
generators it is difficult to see how one may define a vielbein term
in the local supersymmetry theory,
and hence one cannot construct a scalar action which is linear in the
curvatures of the fields. We constructed the group theoretic
curvatures from the algebra and discussed Yang-Mills type actions. We showed that one obtains a rather
complicated quadratic equation of motion if the metric on the base is a function of
the gauge fields or if it is an independent field. The main idea of
the paper is that the twelve dimensional theory should reduce to a
lower dimensional supergravities not by the process of dimensional
reduction but by group theoretic contraction of the geometric theory
underlying the F-algebra. This idea is supported by the fact that one
cannot find suitable compact Lorentzian manifolds of special
holonomy on which to reduce in a supersymmetric fashion.
Upon contraction of the theory we obtain a standard metric and Riemann
tensor. The non-linearity in the Riemann tensor is suppressed by a
factor of the same magnitude as a cosmological constant term. 
Finally, we mention the very important issue of degrees of freedom of
a theory with many times. Assessing correctly the appropriate degrees
of freedom for such problems requires us to understand fully the
difficult question 
generalisation of the Cauchy problem to ultra-hyperbolic
scenarios.

\bigskip
{\bf{Acknowledgements}}

The author would like to thank  Queens' College,
Cambridge and the Department of Applied
Mathematics and Theoretical Physics, Cambridge for their financial and
technical support.


\begin{thebibliography}{10}



\bibitem{12}

M.~Wang, {\em Supergravity theory in twelve dimensions, 
  Proceedings of the fourth {M}arcel {G}rossmann meeting on 
  {G}eneral {R}elativity\/}, pages 1459--1463. Elsevier Science ({\bf 1986}).


D.~Kutasov and E.~Martinec, {\em New principles for string / membrane
  unification\/}, Nucl. Phys. {\bf B477}: 652--674 ({\bf 1996}).

H.~Nishino and E.~Sezgin, {\em Supersymmetric {Y}ang-{M}ills equations in
  (10+2)-dimensions\/}, Phys. Lett. {\bf B388}: 569--576 ({\bf 1996}).




I.~Bars, {\em S theory}, Phys. Rev. {\bf D55}: 2373--2381 ({\bf 1997}).


T.~Hurth, P.~van~Nieuwenhuizen, A.~Waldron and  C.~Preitschopf, {\em
On a possible new $R^2$ theory of supergravity}, Phys. Rev. {\bf D55}:
7593--7614 ({\bf 1997}).


N.~Khviengia, Z.~Khviengia, H.~Lu and C.~N. Pope, {\em Towards a field theory
  of {F}-theory\/}, Class. Quant. Grav. {\bf 15}: 759--773, ({\bf
1998}). 


R.~Manvelian, A. Melikyan and R.~Mkrtchian, {\em Representations and BPS states of
(10+2) superalgebra}, Mod. Phys. Lett. {\bf A13}: 2147--2152 ({\bf 1998}).




H.~Nishino, {\em N=2 chiral supergravity in (10+2)-dimensions as consistent
  background for super(2+2)-brane\/}, Phys. Lett. {\bf B437}:
303--314, ({\bf 1998}).




H.~Nishino, {\em Supersymmetric Yang-Mills theories in $D\geq 12$}, Nucl. Phys. {(\bf B523)}: 440--464, {(1998)}







I.~Rudychev, E.~Sezgin, P.~Sundell, {\em Supersymmetry in dimensions
beyond eleven}, Nucl. Phys. Proc. Suppl. {\bf 68}: 285-0294 ({\bf
1998}).





 R.~Manvelian and R.~Mkrtchian, {\em Towards SO$(10,2)$ invariant
 M-theory: multilagrangian fields},  hep-th/9907011. 




\bibitem{nahm}
W.~Nahm, {\em Supersymmetries and their representations},
Nucl. Phys. {\bf B135}, 149 ({\bf 1978}).



\bibitem{hew1}
S.~F.~Hewson and M.~J.~Perry, {\em The twelve-dimensional super (2+2)-brane\/},
  Nucl. Phys. {\bf B492}: 249--277 ({\bf 1997}).


\bibitem{intersect}
J.~P.~Gauntlett, {\em Intersecting branes\/}, hep-th/9705011.


P.~K.~Townsend, {\em Four lectures on {M}-theory\/}, hep-th/9612121.




\bibitem{vhp}
J.~W.~van~Holten and A.~V.~Proeyen, {\em {N}=1 supersymmetry algebras in {D} =
  2, 3, 4 mod-8\/}, J. Phys. {\bf A15}: 3763 ({\bf 1982}).

\bibitem{hew2}
S.~F.~Hewson. {\em An approach to F-theory}, Nucl. Phys. {\bf B534}:
513--530 ({\bf 1998}).



\bibitem{geometric}

P.~G.~0.~Freund, {\em Introduction to supersymmetry}, Cambridge
University Press.


A.~Chamseddine and P.~C.~West, {\em Supergravity as a gauge theory of
gravity}, Nucl. Phys. {\bf B129} 39--44 ({\bf 1977}).



L.~Castellani, P.~Fr\'{e}, F.~Giani, K.~Pilch and
P.~van~Nieuwenhuizen, Phys. Rev. {\bf D26}: 1481 ({\bf 1982})


R.~d'Auria and P.~Fr\'{e}, Nucl. Phys. {\bf B201}: 101 {(1982)}.

R.~Troncoso and J.~Zanelli, {\em New gauge supergravity in seven and
eleven dimensions}, Phys. Rev.{\bf  D58}: 101703 ({\bf 1998}).

C.~R.~Preitschopf, T.~Hurth, P.~van~Nieuwenhuizen, A.~Waldron, {\em
OSP(1|8) gravity}, Nucl. Phys. Proc. Suppl. {\bf 56B}: 310--317 {(1997)}.



\bibitem{holonomy}
M.~J.~Duff, B.~E.~W.~Nilsson and C.~N.~Pope, {\em Kaluza-Klein
supergravity}. Phys. Rep. {\bf 130}, 1--142 ({bf 1986}).


\bibitem{nolorentz}
H.~Baum and I.~Kath, {\em Parallel spinors and holonomy groups on
pseudo-Riemannian spinmanifolds}, Differential Geometry archive {\tt Math.DG/9803080}







\bibitem{brenner}
See, for example, 
A.~Brenner {\em Poly-parabolic equations}, Mathematical Physics
archive {\tt mp-arc:95-105} (located at 
{\tt http://mpej.unige.ch/mp\_arc/index.html})


\end{thebibliography}
\end{document}